\documentclass[aps,prl,floats,twocolumn,showpacs,floatfix]{revtex4}

\usepackage{graphicx}
\usepackage{textcomp}
\begin{document}
\title{Phonon Excitations of Composite Fermion Landau Levels}
\author{F. Schulze-Wischeler$^{1}$, F. Hohls$^{1}$, U. Zeitler$^{1,2}$, D. Reuter$^{3}$, A. D.
Wieck$^{3}$, and R. J. Haug$^{1}$
  \vspace{1mm}}
\affiliation{$^{1}$Institut f\"ur Festk\"orperphysik, Universit\"at Hannover,
  Appelstra\ss{}e 2, 30167 Hannover, Germany \\
$^{2}$High Field Magnet Laboratory, NSRIM, University of
Nijmegen, Toernooiveld 7, 6525 ED Nijmegen, The Netherlands\\
$^{3}$Lehrstuhl f\"ur Angewandte Festk\"orperphysik,
Ruhr-Universit\"at Bochum, Universit\"atsstra\ss{}e 150, 44780
Bochum, Germany \vspace{3mm}}
\date{February 26, 2004}
\begin{abstract}
Phonon excitations of fractional quantum Hall states at filling
factors $\nu = \frac{1}{3}, \frac{2}{5}, \frac{4}{7}, \frac{3}{5},
\frac{4}{3},$ and $\frac{5}{3}$ are experimentally shown to be
based on Landau level transitions of Composite Fermions. At
filling factor $\nu = \frac{2}{3}$, however, a linear field
dependence of the excitation energy in the high-field regime
rather hints towards a spin transition excited by the phonons. We
propose to explain this surprising  observation by an only
\emph{partially} polarized $\frac{2}{3}$-ground-state making the
energetically lower lying spin transition also allowed for phonon
excitations.

\end{abstract}
\maketitle
\newcommand{\muB}{\mu_{\tiny{\mathrm{B}}}^{}}
\newcommand{\rhoxx}{\rho_{\mathrm{xx}}^{}}
\newcommand{\rhoxy}{\rho_{\mathrm{xy}}^{}}
\newcommand{\Rpkt}{R_{\mathrm{2point}}}

The Coulomb interaction in a two-dimensional electron system
(2DES) subjected to a quantizing magnetic field leads to the
formation of new, fractionally charged quasi-particles at Landau
level filling factors $\nu = p/q$ ($q$ is an
odd-integer)~\cite{FQHE1,FQHE2}. In the last decade this
fractional quantum Hall (FQH) effect was very effectively
described in the framework of Composite Fermions (CFs)
\cite{Jain89}. At a fractional filling factor with {\sl even}
denominator, $\nu= 1/2m$, these quasiparticles are formed by
attaching an even number $2m$ of flux quanta $\phi_0$ to each
electron, i.e. two flux quanta at $\nu = 1/2$. Their effective
mass $m^*$ is originating from the Coulomb
interaction~\cite{HLR93}.

The ground state of a 2DES at a fractional filling factor $\nu =
p/q$ is a collective wave function~\cite{FQHE2} with finite
wave-vector collective excitations~\cite{GMP,KWJ} directly
accessible by e.g. Raman techniques~\cite{Pinczuk88,Kang00},
photoluminescence~\cite{Kukushkin99} or phonon absorption
experiments~\cite{Mellor95,Zeitler99}. In a simple picture these
excitations originate from the level structure of CFs in an
effective magnetic field, $B_{eff} = B - B\;(\nu = \frac{1}{2})$,
with an effective (integer) filling factor, $p =
\frac{\nu}{1-2\nu}$~\cite{Jain89}. The levels can then be
described as spin-split Landau levels of CFs and, therefore,
excitations can be interpreted as either Landau level transitions
or spin-excitations.

In this Letter we use phonons to probe the FQH excitation spectrum
at filling factors $\nu$ = 1/3, 2/5, 4/7, 3/5, 2/3, 4/3, and 5/3
when varying the electron densities in the same sample over a wide
range. At a given filling factor nearly all gaps measured show a
square-root dependence on the magnetic field, strongly suggesting
that we probe Landau level transitions of CFs. All these data are
described by one single fit parameter related to the CF effective
mass. Surprisingly, the gap at $\nu = 2/3$ displays a {\sl linear}
field dependence, rather related to a spin transition forbidden
for phonon excitations. This observation is a clear hint that the
2/3-ground-state is not fully spin-polarized in high $B$-fields
and that a strict separation between spin transitions and CF
Landau-level transitions is no more possible.

Our sample consists of a high-mobility AlGaAs/GaAs-heterojunction
grown on a 2~mm thick GaAs wafer. On the front side containing the
2DES we patterned a $w = 90~\mu$m wide meander extending over a
total length $l = 10$~mm on an area $A = 1\times1$~mm$^2$. The
huge aspect ratio $l/w = 111$ maximizes the $\rhoxx$-contribution
to the two-terminal resistance and thereby allows to measure
smallest changes in $\rhoxx$. The Ohmic contacts to the 2DES are
placed at the edges of the sample far away from the meander to
avoid any phonon interactions with the contacts. We took great
care that the contact resistances ($ <10~\Omega$) only play a
negligible role in the measured two-terminal resistance. Using the
persistent photoconductivity we varied the electron concentration
in several steps from $n = 0.89\times 10^{15}$~m$^{-2}$ (mobility
$\mu = 102$~m$^2$/Vs) to $n = 1.50\times 10^{15}$~m$^{-2}$
(mobility $\mu = 193$~m$^2$/Vs).

The sample is mounted on the tail of a dilution refrigerator in a
superconducting magnet with maximum fields up to 13~T and
connected to high frequency coaxial cables. Great care was taken
to assure a proper thermal anchoring of the cables. We achieve a
2DES temperature $T_{\mathrm{2DES}} \lesssim 100$~mK for a
cryostat base temperature of 75~mK.

\begin{figure}[t]  
  \begin{center}
  \resizebox{0.9\linewidth}{!}{\includegraphics{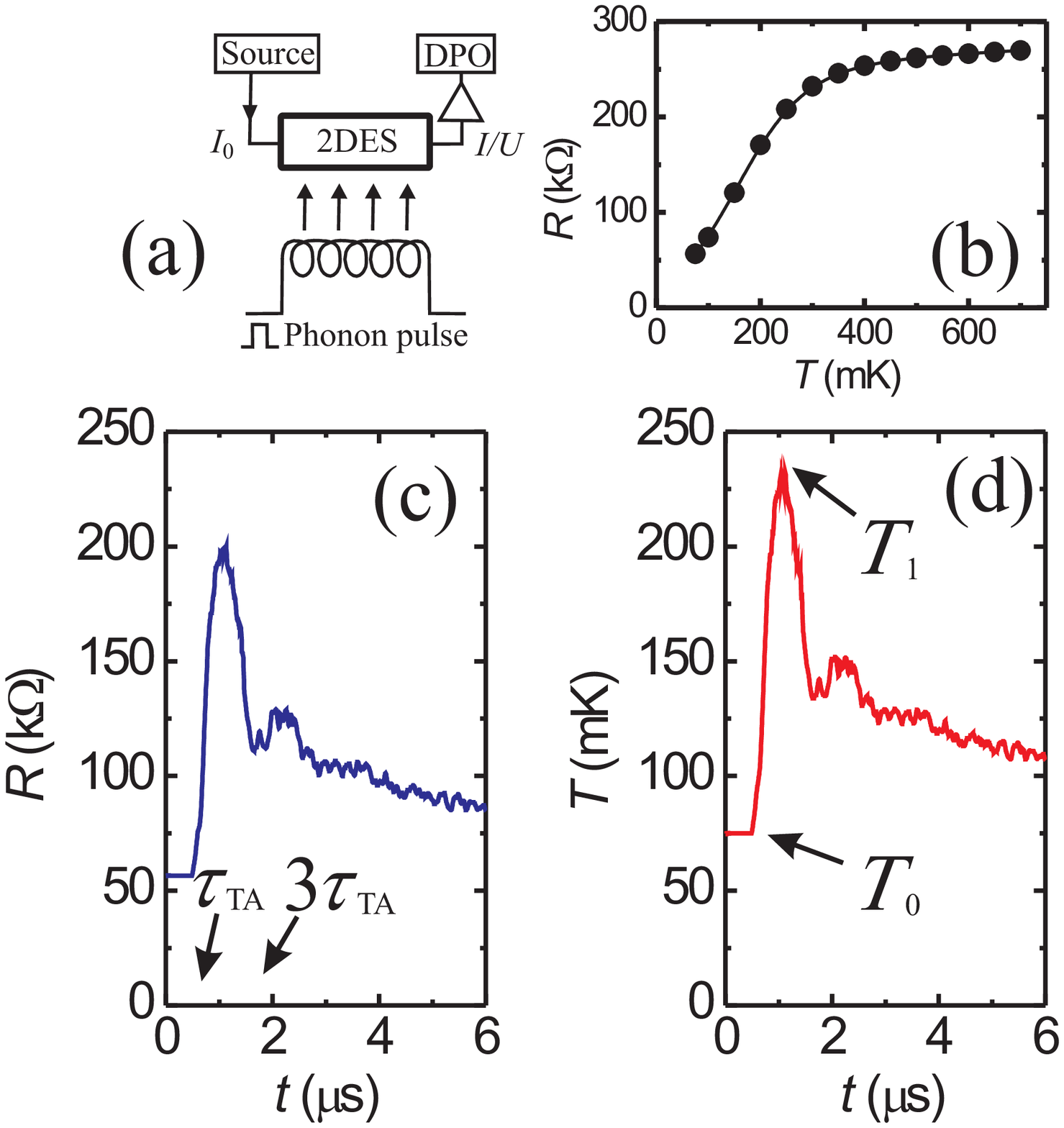}}
  \end{center}
\caption{(color online). a) Schematic experimental setup with the
phonon emitter on the back side and the 2DES at the front. Time
resolved readout is achieved by an ultra fast digital phosphore
oscilloscope (DPO).
\newline (b) Steady-state calibration of the 2DES resistance at
$\nu = 2/3$ for $n = 0.89 \times 10^{15}$~m$^{-2}$.\newline (c)
Phonon signal for the same $\nu$ and $n$ after emitting
$T_{\mathrm H} = 2.09$~K phonons during $\tau_{\mathrm H} = 10$~ns
at $t = 0$. The 2DES was excited from a base temperature $T_0 =
75$~mK. \newline (d) 2DES temperature versus time as deduced from
raw phonon signal curve (c) using the calibration (b).  }
  \label{Fig-PS-Signal}
\end{figure} 

The experimental setup is shown schematically in
Fig.~\ref{Fig-PS-Signal}a: A thin constantan film acting as a
phonon emitter is placed on the polished back side of the sample.
By passing a short current pulse during a time $\tau_{\mathrm H}$
through this heater, non-equilibrium phonon pulses are created at
the heater-GaAs interface. They are characterized by a black-body
spectrum at a temperature $T_{\mathrm H} = (P_{\mathrm H}/\sigma
A_{\mathrm H} + T_{0}^{4})^{1/4}$, where $A_{\mathrm H}$ is the
heater area, $P_{\mathrm H}$ is the power dissipated in the
heater, $\sigma = 524$~W/m$^2$~K$^4$ is the acoustic mismatch
constant between constantan and GaAs, and $T_0$ is the GaAs
lattice temperature \cite{Kent96}. When entering the GaAs, the
non-equilibrium phonons travel ballistically through the
2~mm-thick substrate. After a time of flight $\tau_{\mathrm{LA}} =
0.42~\mu$s and $\tau_{\mathrm{TA}} = 0.6~\mu$s for longitudinal
acoustic (LA) phonons and  for transverse acoustic (TA) phonons,
respectively, they hit the 2DES and a small part of their energy
is absorbed \cite{Schuwi02}. As a consequence, the 2DES
temperature increases, directly measured by a change of its
resistance. This resistance change is detected by the current
change at constant voltage using a 10-MHz current amplifier and a
digital phosphor oscilloscope to average over up to a few million
traces.

In Fig.~\ref{Fig-PS-Signal}c we have plotted a characteristic
signal at filling factor $\nu = 2/3$. Due to their strong
focussing~\cite{Zeitler99} mainly TA phonons are visible, with a
first peak starting around $\tau_{\mathrm{TA}} = 0.6~\mu$s and a
second due to multiply reflected phonons after
$3\tau_{\mathrm{TA}}$. After typically 1 ms (depending on the
total power dissipated inside the GaAs substrate), the sample has
cooled down back to its base temperature and the experiment is
repeated a few million times.

By using the temperature dependent resistance measured under
equilibrium conditions (as plotted in Fig.~\ref{Fig-PS-Signal}b)
the raw phonon signal curve can be translated into a 2DES
temperature versus time. The reliability of this procedure is
checked as follows: After a certain time ($\approx 10~\mu$s), all
non-equilibrium phonons induced with the heater are thermalized in
the GaAs substrate and the 2DES and the substrate are in thermal
equilibrium. This is experimentally measured by a merely changing
2DES temperature. This measured temperature agrees well with the
theoretically expected one as deduced from the total energy
dissipated in the heater, $P_{\mathrm{H}} \tau_{\mathrm H}$, and
the specific heat of the GaAs substrate.

In order to extract quantitative data from our experiments we use
a simple model to describe the phonon absorption in the 2DES. In
the most general case, the differential temperature gain $dT$ of
the 2DES within a time interval $dt$ is given by
\begin{equation}\label{model1}
  C(T) \, dT = r(T,T_{\mathrm H})P_{\mathrm H} \, dt - P_{\mathrm
  e}(T,T_0) \, dt
\end{equation}
, where $C(T)$ is the 2DES's specific heat, $r(T,T_{\mathrm H})
P_{\mathrm H}$ is the phonon energy absorbed by the 2DES with an
absorption coefficient $r$ depending on the heater temperature
$T_{\mathrm H}$ and the 2DES temperature $T$, and $P_{\mathrm
e}(T,T_0)$ is the energy emitted by the 2DES, depending on $T$ and
the equilibrium substrate temperature $T_0$. In our experiments we
use very short (10~ns) heater pulses with a moderate heater power
$P_{\mathrm H}$. Consequently, the peak height of the first
ballistic phonon signal peak is dominated by absorption and the
emission term can be ignored on these short time scales.

In a first set of experiments, we calibrate the relative specific
heat of the 2DES at given fractional filling factor. The maximum
2DES temperature on the first ballistic phonon peak, $T_1$ (see
Fig.~\ref{Fig-PS-Signal}d), is measured as a function of the 2DES
base temperature $T_0$, with a fixed duration and a constant
amplitude of the heater pulse. Since all the $T_0$ used are
distinctively lower than the energy gaps at these filling factors
we always deal with a situation where the quasiparticle ground
states are almost full and their excitations are almost empty. As
a consequence, the relative proportion of phonons absorbed is
independent from $T_0$, and we can approximate $r(T,T_{\mathrm
H})\rightarrow r_0$. Integrating Eq.~(\ref{model1}) over the pulse
length with these assumptions, we get:
\begin{equation}\label{model2}
  \int_{T_0}^{T1}C(T) \, dT =  r_0 P_{\mathrm H}\tau_{\mathrm H}
\end{equation}
Using the mean value theorem for this integral equation we can
determine the relative specific heat $C(T)/r_0$ of the 2DES from
our set of experiments where we measured $T_1$ for fixed heater
power and varying $T_0$. As a consistency check we performed the
same set of experiments with different heater powers and find a
very comparable temperature dependence of $C(T)$.

\begin{figure}[t]  
  \begin{center}
  \resizebox{0.95\linewidth}{!}{\rotatebox{270}{\includegraphics{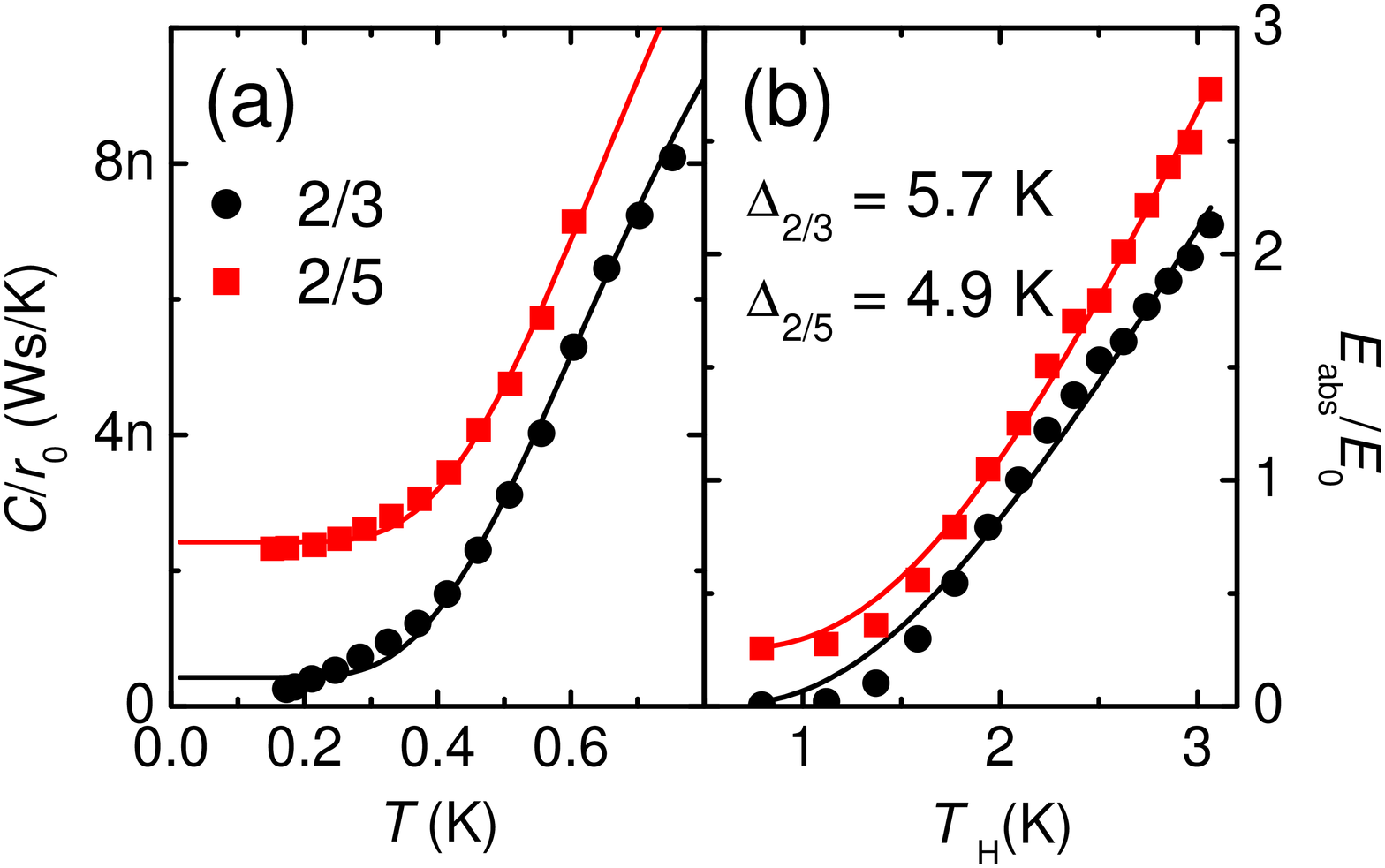}}}
  \end{center}
\caption{(color online). (a) Examples of the relative specific
heat at filling factors $\nu$ = 2/3 and 2/5 measured by phonon
absorption for sample \#5  ($n = 1.21 \times 10^{15}$~m$^{-2}$).
Data for 2/5 are shifted for clarity by 2n. The lines are fits
according to Ref.~\cite{Chakraborty97} plus a constant.
\newline
(b) Relative phonon energy absorbed by the 2DES, normalized to its
value $E_0$ for $T_{\mathrm H} = 2.09$~K, for a 10~ns pulse at
filling factors $\nu$ = 2/3 and 2/5 (Data for 2/5 are shifted by
0.25 for clarity). The curves are fits for an excitation across a
gap $\Delta_{2/3} = 5.7$~K and $\Delta_{2/5} = 4.9$~K.}
\label{FigCalib}
\end{figure}

In Fig.~\ref{FigCalib}a the measured relative specific heats,
$C(T)/r_0$, are shown exemplarily for filling factors $\nu = $ 2/3
and 2/5. We should note that we cannot determine the absolute
value of $C(T)$ due to the unknown absorption coefficient $r_0$
($0<r_0<1$). The lines shown in Fig.~\ref{FigCalib}a are fits to
the theoretical predictions for the specific heat of a 2DES
$C_{\mathrm{2DES}} \propto \frac{1}{T^2}e^{{-\Delta_{\mathrm
C}/T}}$ \cite{Chakraborty97} plus a small empirical constant,
taking into account an additional contribution to $C(T)$ possibly
resulting from a finite (thermodynamic) density of states inside
the excitation gap~\cite{Zeitler99}.

In a second set of phonon absorption experiments we can now
determine the energy gaps at fractional filling factors. This
time, the heater temperature $T_{\mathrm H}$ is varied for a fixed
base temperature $T_0$. By increasing $T_{\mathrm H}$ the number
of phonons for every wavelength is increased. Since the major
contribution to the phonon absorption is predominantly due to
excitations around a gap $\Delta$, the total energy absorbed by
the 2DES increases as
\begin{equation}
\label{abs}
 E_{\mathrm{abs}} \propto \frac{1}{\exp(\Delta /T_{\mathrm H}) - 1}
\end{equation}

The absorbed phonon energy, $E_{\mathrm{abs}} = r(T,T_{\mathrm H}
) P_{\mathrm H} \tau_{\mathrm H}$, as a function of $T_{\mathrm
H}$ is deduced from the measured $T_1$ and $T_0$ by integrating
Eq.~(\ref{model1}) using the previously determined specific heat.
Again, the emission term is neglected for the short time scales
considered. In Fig.~\ref{FigCalib}b we show results for filling
factors $\nu = $ 2/3 and 2/5. The gap values are obtained by
averaging over several experiments using different heater powers
and base temperatures. All the individual gaps measured are within
$\pm10\%$ of the average value. The solid lines in
Fig.~\ref{FigCalib}b show fits using Eq.~(\ref{abs}) and indeed,
the experimental data match a model
with phonon excitation across a single energy gap $\Delta$.\\

\begin{figure}[t]  
  \begin{center}
  \resizebox{0.9\linewidth}{!}{\rotatebox{270}{\includegraphics{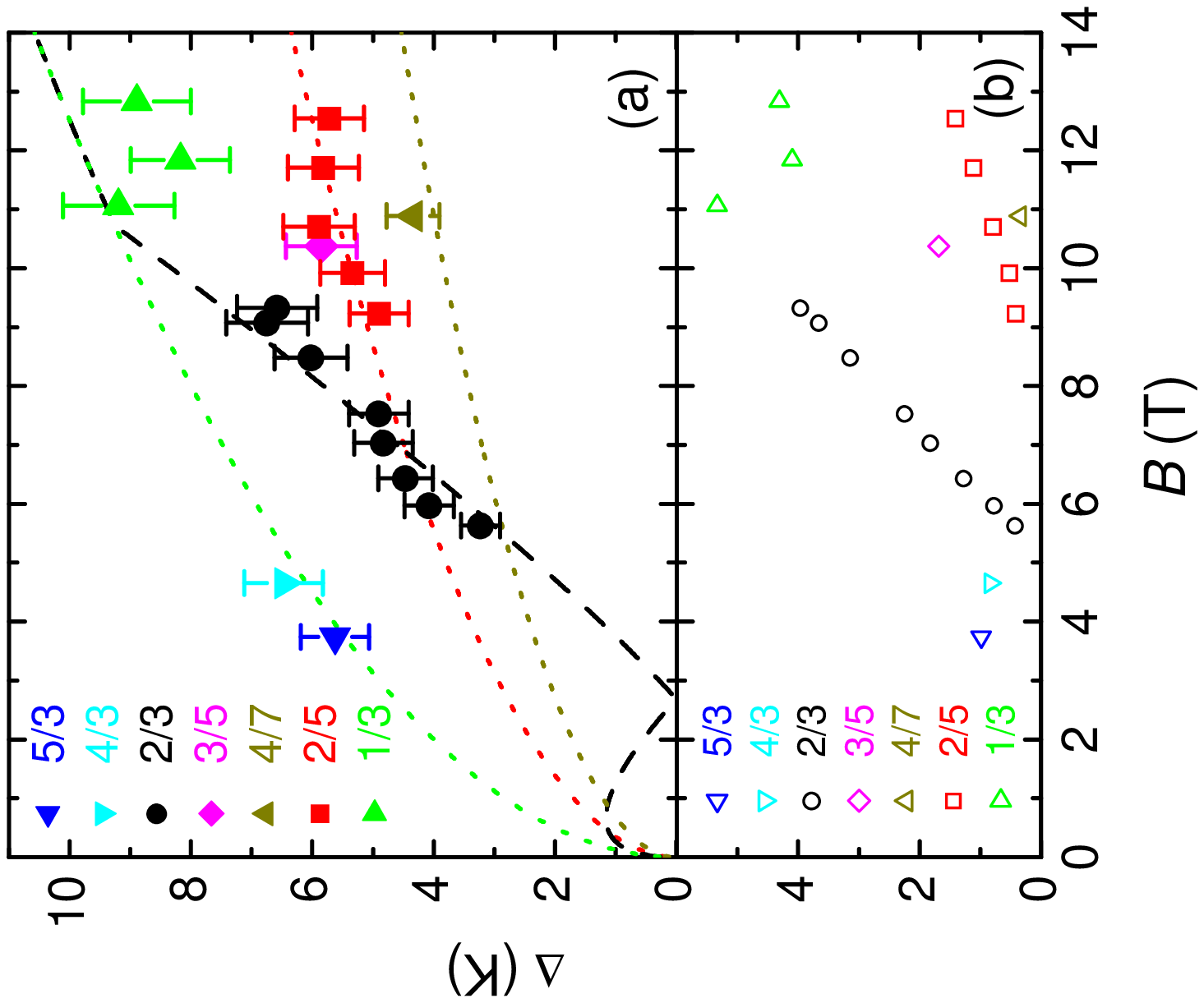}}}
  \end{center}
\caption{(color online). (a) Compilation of all phonon excitation
gaps $\Delta$ measured at various filling factors as a function of
the magnetic field (i.e. the electron concentration for each
filling factor). The dotted lines are CF Landau level transitions
adjusted to our phonon data with one single mass parameter $\alpha
= 0.158~(\pm 0.006)$. The dashed line represents the expected form
of a normally forbidden spin transition at $\nu = 2/3$.
\newline (b) The transport gaps are systematically lower then the
phonon excitation gaps in (a).} \label{FigGaps}
\end{figure} 
Using such an elaborated series of calibrations we can now
investigate in detail the excitation gaps. As listed in Table~I,
we have measured the gaps for eight different electron
concentrations at filling factors $\nu = $ 2/3, 2/5, and 1/3 up to
the highest magnetic fields accessible in our magnet. These
filling factors all showed well developed minima for all the
concentrations used. Additional filling factors $\nu = $ 4/3, 5/3,
3/5 and 2/7 are only clearly pronounced for the highest electron
concentration and, accordingly, phonon gaps were only measured in
sample \#8 for these filling fractions.

All gaps measured at various filling factors and electron
concentrations are compiled in  Fig.~\ref{FigGaps}a. For
comparison, we also show the energy gaps deduced from activated
transport measurements~\cite{schuwiPRL} in Fig.~\ref{FigGaps}b. In
order to discuss these results for different filling factors $\nu$
we describe them within the CF picture \cite{Jain89}, where the
FQH filling factors $\nu = \frac{p}{2p+1}$ are mapped to integer
CF filling factors $p$. The energy levels for $|p|$ filled CF
Landau levels can then be written \cite{Mariani02} as:
\begin{equation}
\label{Landau Level}
  E_{ns} (p) = (n+\frac{1}{2})\hbar \omega^*_{\mathrm{c}}(p) + s
  g^* \mu_{\mathrm B} B
\end{equation}
Here the CF cyclotron energy is given purely by the Coulomb
interaction and thus follows the form $\hbar \omega^*_{\mathrm c}
= \hbar e B/(m^* (2p \pm 1))$ with a CF mass $m^* = m_e  \alpha
\sqrt{B[\mathrm T]}$ \cite{Park98}.

Since phonons carry no spin we expect that, in a phonon absorption
experiment, the lowest lying excitations of a CF state $p$ are
Landau level transitions from $n$ to $n$+1 with the same spin $s$.
The corresponding energy gap $\hbar \omega^*_{\mathrm{c}}(p)$ can
now be adjusted to our data by  one single dimensionless mass
parameter $\alpha$. In Fig.~\ref{FigGaps} the fits of such Landau
level transitions to all the data at $\nu = $ 1/3, 2/5, 4/7, 3/5,
4/3, and 5/3 are shown yielding $\alpha = 0.158~(\pm 0.006)$
(dotted lines). Here, $\nu = 4/3 = 1 +1/3$ and $5/3 = 1 + 2/3$ are
treated as 1/3 respectively 2/3 plus one inert fully occupied
Landau level. The experimentally determined CF mass parameter
$\alpha$ is in astonishing agreement with the theoretical
predictions in Eq.~(1) of Ref.~\onlinecite{Park98}.

\begin{table}[t]
\caption{Electron concentrations, mobilities and energy gaps
measured by phonon absorption at filling factors 2/3, 2/5, and
1/3. When no values are given, the field required was too large to
be accessed in our magnet.\\} \label{table1}
 \centering
\begin{tabular}{|c|cc|ccc|}  \hline\hline
sample &~~~~~~~$n$~~~~~~~&~~~~~~~$\mu$~~~~~~~& \multicolumn{3}{c|}{ $\Delta$ [K] for $\nu = $ }\\
   \#  &[10$^{15}$ m$^{-2}$]&[m$^2$/Vs]
   &~~~~~$\frac{2}{3}$~~~~~&~~~~~$\frac{2}{5}$~~~~~&~~~~~$\frac{1}{3}$~~~~~\\[0.5em]
   \hline
   1 &0.89       &102                  &3.2         &4.9        &9.2\\
   2 &0.95       &109                  &4.1         &5.3        &8.2\\
   3 &1.03       &119                  &4.5         &5.9        &8.9\\
   4 &1.13       &131                  &4.8         &5.8        &---\\
   5 &1.21       &144                  &4.9         &5.7        &---\\
   6 &1.36       &168                  &6.0         &---        &---\\
   7 &1.46       &187                  &6.7         &---        &---\\
   8 &1.50       &193                  &6.6         &---        &---\\[0.5em]
   \hline \hline
\end{tabular}
\end{table}

Compared to all phonon gaps measured at the above mentioned
filling factors, the phonon absorption data at filling factor 2/3,
are distinctly different: The measured excitation gaps can in no
way be described with the square-root dependence of CF Landau
level excitations. They rather show a linear dependence,
$\Delta_{2/3}\propto (B-B_p)$, strongly suggesting that they are
related to a spin gap, which is normally not directly accessible
by phonon excitations. Here $B_p \approx 2.8$~T is the field where
the 2/3 state changes from a spin-unpolarized to a spin-polarized
state. Indeed, when we quantitatively compare the expected field
dependence of the spin gap with our data we find a remarkable
agreement (dashed line in Fig.~\ref{FigGaps}a). This
interpretation is also supported by the transport gaps measured at
$\nu = 2/3$ which have the same linear behavior, reduced by a
constant due to disorder as seen in Fig.~\ref{FigGaps}b. All other
transport gaps are also systematically lower than the phonon
excitation gaps due to disorder effects \emph{and} because
temperature can also couple to spin flip excitations.

The fact that we observe spin-related excitation gaps with phonons
indicates that the $\nu = 2/3$ state can not be described with
independent spin and Landau level indexes. This supposition is
also supported by recent theoretical \cite{Apalkov01} and
experimental \cite{Freytag01} evidence suggesting that $\nu = 2/3$
state is not fully polarized, even in high magnetic fields. As a
result, we may speculate that the complexity of the 2/3 state is
responsible for the appearance of a spin-forbidden transition in
the phonon absorbtion.

In conclusion, we have measured phonon excitation gaps in the FQH
regime for filling factors $\nu = $ 1/3, 2/5, 4/7, 3/5, 2/3, 4/3,
and 5/3 for eight different electron densities. For all filling
factors besides $\nu = 2/3$ the measured gaps can be well
described in the framework of Landau level transitions of CF
involving no spin flip. The gaps measured at $\nu = 2/3$, however,
correspond to a normally forbidden spin transition, pointing
towards a complex not fully polarized ground state.

We acknowledge financial support by BMBF and DFG priority program
``quantum Hall systems''.


\end{document}